# Effect of Chiral Damping on the dynamics of chiral domain walls and skyrmions

C.K. Safeer[1,2,3], Mohamed-Ali Nsibi[1], Jayshankar Nath[1], Mihai Sebastian Gabor[4], Haozhe Yang[1], Isabelle Joumard[1], Stephane Auffret[1], Gilles Gaudin[1], & Ioan-Mihai Miron[1]*

[1]*Univ. Grenoble Alpes CNRS, CEA, Grenoble INP, SPINTEC, Grenoble, France*
[2]*CIC nanoGUNE BRTA, 20018 Donostia-San Sebastian, Basque Country, Spain.*
[3]*Department of Physics, Clarendon Laboratory, University of Oxford, Oxford, United Kingdom*
[4]*C4S, Physics and Chemistry Department, Technical University of Cluj-Napoca, Cluj-Napoca, Romania*

\* To whom correspondence should be addressed.
E-mail: mihai.miron@cea.fr

**Abstract**

Friction plays an essential role in most physical processes that we experience in our everyday life. Examples range from our ability to walk or swim, to setting boundaries of speed and fuel efficiency of moving vehicles. In magnetic systems, the displacement of chiral domain walls (DW) and skyrmions (SK) by Spin Orbit Torques (SOT), is also prone to friction. Chiral damping ($\alpha_c$), the dissipative counterpart of the Dzyaloshinskii Moriya Interaction (DMI), plays a central role in these dynamics. Despite experimental observation, and numerous theoretical studies confirming its existence, the influence of chiral damping on DW and SK dynamics has remained elusive due to the difficulty of discriminating from DMI. Here we unveil the effect that $\alpha_c$ has on the flow motion of DWs and SKs driven by current and magnetic field. We use a static in-plane field to lift the chiral degeneracy. As the in-plane field is increased, the chiral asymmetry changes sign. When considered separately, neither DMI nor $\alpha_c$ can explain the sign reversal of the asymmetry, which we prove to be the result of their competing effects. Finally, numerical modelling unveils the non-linear nature of chiral dissipation and its critical role for the stabilization of moving SKs.

# Introduction

The observation of DMI[1,2] in metallic multilayers has allowed significant advances in understanding the mechanism of the current induced DW motion[3–6] as well as the development of the current induced SK dynamics[7–14]. These phenomena, potentially useful for applications are closely related. Their efficiency relies on two pillars: efficient SOT and strong DMI. In materials with broken inversion symmetry and with large spin orbit interaction, the electric current produces a damping like torque (DL)-SOT[15,16], and the DMI effective field, deriving from a chiral energy contribution, imposes a Néel DW structure[4]. In this configuration, the magnetization of the DWs is parallel to the current (Figure 1a), which makes the SOT most efficient in producing DW displacements.

By analogy with DMI, which is a chirality dependent energy, the chiral damping is a chirality dependent dissipation. Several theoretical approaches have successfully predicted different possible origins, as well as diverse manifestations[17–22]. The main feature though, is that in practice, as the chirality is modified, the damping coefficient changes. A simple way to express this for the case of a DW in a perpendicularly magnetized material of interest here, is $\alpha_c = \alpha_1 + \alpha_2(\vec{m}_{ip} \cdot \vec{\nabla}\vec{m}_z)$. Here $\vec{m}_{ip} \cdot \vec{\nabla}\vec{m}_z$ is the projection of the in-plane DW magnetization ($\vec{m}_{ip}$) on the normal to the DW position ($\vec{\nabla}\vec{m}_z$) (Figure 1a, 1c). Because this projection can be negative, $\alpha_1$ must be larger than $\alpha_2$ to ensure that the damping always remains positive.

Having the same symmetry and possibly driven by the same microscopic interactions, the effects of $\alpha_c$ cannot be easily separated from those of DMI. The ability to isolate them experimentally is crucial for the fundamental understanding of chiral phenomena in magnetism[23], as well as from the purely pragmatic perspective of applications. Dissipative processes are fundamental to enable high performance, energy efficient, computing and storage devices based on magnetic SKs and chiral DWs[24–26].

In this work, we separate the respective contributions of DMI and $\alpha_c$ on the dynamics of DWs and skyrmionic bubbles and prove that the two phenomena compete (Figure 1a). In particular, $\alpha_c$ modulates the DW and SK velocity and modifies the shape of the SK bubbles during their motion (Figure 1b), affecting their dynamical stability[13,27].

# Results and discussion

As a material platform for our study, we employ a $Pt_{3\ nm}/Co_{0.6\ nm}/Pt_{1.5\ nm}$ tri-layer. The first reason is that in this material we have already observed evidence of chiral damping[17]. The second reason is that DMI is relatively weak, and allows controlling the chirality by moderate in-plane magnetic fields ($H_{ip}$).

We study the chiral dynamics in the flow regime, by measuring the displacements of DWs and SKs induced by nanosecond current or out-of-plane field pulses (see methods) using wide field Kerr microscopy. Our study covers all possible types of chiral dynamics: field induced DW motion (FIDM), current induced DW motion (CIDM) and current induced SK motion (CISKM). These dynamics are very different: FIDM can be either turbulent or steady motion, the CIDM has a single highly non-linear regime, while the CISKM includes 2D effects, not present in the case of CIDM. The chiral effects are extracted by monitoring the asymmetric motion of DWs with opposite polarity (up/down and down/up). When applying a sufficiently strong $H_{ip}$, as the magnetization of both DWs aligns with the field, their chirality becomes opposite (Figure 1a). In this situation, the two DWs will have different energy (due to DMI) and different damping (due to $\alpha_c$). This reveals the competing effects of DMI and $\alpha_c$ on the DW motion asymmetry in all experiments.

Our analysis is supported by a numerical model, based on a $q$-$\phi$ approach, (see supplementary information S1). The physical parameters used in the simulations are either measured independently or obtained from fitting the experimental data. Because the experimental features outnumber the free variables in the model, the parameters are uniquely determined (S1). The simulations reproduce accurately the ensemble of our experimental results, by using the same set of values for all cases. Micromagnetic simulations based on MuMax[34] (Supplementary information S10) are further used to assess the influence of $\alpha_c$ on the SK stability.

For the sake of clarity, in sections I, II, and III we present separately the three different experiments. In section IV, we present theoretical predictions of the model, using parameter values required for applications.

**I) Current Induced DW motion**

The samples are patterned in the form of 10 μm long and 1 μm wide wires. The DWs are prepared using a perpendicular magnetic field. Under the effect of ns current pulses, the DWs shift in the direction of the current, indicating a left-handed DW chirality[4]. The DW velocity increases rapidly above $1 \cdot 10^8$ A/cm$^2$ (Figure 2c). This rapid increase characteristic of the thermally activated DW motion stops above $2 \cdot 10^8$ A/cm$^2$, signaling the limit of the flow regime. The saturation of the velocity observed above $2 \cdot 10^8$ A/cm$^2$ is characteristic of the SOT-DMI mechanism in the flow regime, typically observed for materials with weak DMI[4].

To evidence directly the chiral effects, we compare the displacements of up/down and down/up DWs as we apply $H_{ip}$ (See Figure S7 for the detailed Kerr images). The current density is fixed at $1 \cdot 10^8$ A/cm$^2$, sufficient to induce DW motion in the absence of $H_{ip}$, but not to provoke nucleations at the largest $H_{ip}$. Without $H_{ip}$, both DWs move in the same direction, shifting the domain that they enclose (Figure 2a) along the current flow. As $H_{ip}$ is increased, the two DWs in the wire no longer move with the same velocity, leading to a contraction of the domain (Figure 2a; 2b). This is because $H_{ip}$ magnetizes them in the same direction. According to the DMI-SOT model it is expected that, as $H_{ip}$ becomes much larger than $H_{DMI}$, their motion should be symmetric[5,6], without any net displacement of the central domain (Figure 2b). Experimentally, the DWs indeed move symmetrically, but only at fields of approximately 80 mT. Above this value, the slow moving DW (blue arrow) becomes faster (Figure 2a; 2b), shifting the enclosed domain against the electric current. This is to say the unidirectional component of the motion has been reversed.

Though the DW motion at $H_{ip}$ =0 is hindered by pinning, the reversed unidirectional motion is observed at large velocity (> 20 m/s) in the flow regime. Therefore, we can safely test the reversal of the unidirectional component of motion within our model. This result can only be reproduced by including both DMI and $\alpha_c$ (light blue and red arrows in Figure 2b). In the absence of DMI there is no DW motion at $H_{ip}$=0 (Figure S5), and without $\alpha_c$, the DW motion at large $H_{ip}$ is fully symmetric (orange and green arrows in Figure 2b). The reason for this dependence is that at large $H_{ip}$, the influence of $H_{DMI}$ becomes negligible and the magnetization of the two DWs has the same orientation. Therefore, the effective field produced by SOT on the two DWs is identical. The reversal of the velocity asymmetry proves that DWs with opposite chirality have different damping. Thus, $\alpha_c$ is required to explain quantitatively and qualitatively the DW dynamics.

## II) Field induced DW motion

We start by measuring the DW velocity vs. the out-of-plane magnetic field ($H_Z$) without applying $H_{ip}$ (Figure 3a). We prepare bubble domains in an un-patterned film by applying a small $H_Z$ field. To reach the flow DW motion and avoid the influence of pinning, we apply strong ( up to -217 mT) but short (40 ns) $H_Z$ pulses. The DW velocity (Figure 3a) increases linearly from 40 m/s to 70 m/s when $H_Z$ is larger than -110 mT, confirming the observation of the flow regime. For weaker $H_Z$ values, the velocity drops fast, indicating a significant DW pinning below 40 m/s. In agreement with the low DW mobility, the numerical fitting of this curve reveals that the motion of the DW is turbulent. This is consistent with the low Walker breakdown field characteristic for materials with weak DMI[4].

We then measure the asymmetry induced by $H_{ip}$ [17,28–30] on the velocity of up/down and down/up DWs (See Figure S8 for the detailed Kerr images). The effective in-plane field will be $H_{ip}$+$H_{DMI}$ for down/up DWs, and $H_{ip}$-$H_{DMI}$ for up/down DWs. Therefore, the presence of DMI is recognized by a lateral shift of the velocity vs $H_{ip}$ curves. To a first approximation, $H_{DMI}$ can be extracted directly from the magnitude of the lateral shift[29,30]. The measured DW velocity (Figure 3b) exhibits an almost parabolic dependence on $H_{ip}$. The small lateral shifting of the curves indicates a small $H_{DMI}$. Both the sign and magnitude of $H_{DMI}$ are consistent with the CIDM experiment presented above. However, at large $H_{ip}$ (> 150 mT) there is a deviation from this behavior: as the velocity curves for the up/down (blue) and down/up (red) DWs cross, the DW motion asymmetry reverses. The fact that the asymmetry reversal occurs at large velocity ( > 100 m/s), where the DW motion is in the flow regime, allows us to apply our model to reproduce this experimental feature.

The simulations show that at low $H_{ip}$, where DW motion is turbulent, because of the periodic changes of the chirality, $\alpha_c$ has little effect on the DW velocity. As $H_{ip}$ exceeds a critical value required to stabilize the DW magnetization the DW periodic transformations cease. In this second regime of steady motion, the effect of chiral damping on the asymmetry is stronger than the opposing effect of DMI, leading to the change in sign. Indeed, if $\alpha_1$=0.4 and $\alpha_1$=0, in the steady regime the curves approach asymptotically but the crossing never occurs.

A simple way to verify this scenario experimentally is to repeat the measurements at weaker $H_z$. In this case, a smaller $H_{ip}$ should be sufficient to restore the steady DW motion and achieve the asymmetry reversal. Indeed, the velocity curves measured at $H_Z$= -55 mT exhibit a reversed asymmetry over an extended interval (Figure 3c), starting at $H_{ip}$= 40 mT, while the positive asymmetry region is nearly suppressed. Once again, the model reproduces this tendency (Figure 3f). Because the DW motion at $H_Z$= -55 mT starts to be affected by pinning, which is not included in the model, a better match between model and experiment is obtained using $H_Z$= -10 mT in the calculation (inset of Figure 3f). This confirms that the reversal of the velocity asymmetry is caused by $\alpha_c$ and occurs as the DW motion becomes steady at large $H_{ip}$.

**III) Current induced bubble deformation**

Our next step is to evidence the effect of chiral damping on the current induced motion of the SK bubbles. For these experiments, the layers are patterned into 10 µm wide wires, large enough to contain the displacements of a single bubble (Figure 4a). The bubbles can undergo plastic deformations, as they are not stabilized by the dipolar energy or DMI, but by local pinning. In this sense, they differ from stable SKs, but they have an important advantage: the shape resulting from the deformation is stable over time. Therefore, it is possible to measure the current induced distortion of the bubbles. Such effects cannot be evidenced in stable SKs, which would return to their equilibrium position after the current is removed.

The experiments involving SKs also exhibit the features that we have observed in the CIDM experiments present above: *i)* bubbles are initially shifted in the direction of the current; *ii)* as $H_{ip}$ increases, the bubble growth (shrinking) becomes predominant over the shifting; *iii)* at the largest $H_{ip}$ ( > 80 mT) the shift is reversed, and the bubbles displace in the opposite direction.

The 2D geometry employed here allows to evidence two additional features. First, at zero $H_{ip}$ the motion is asymmetric with respect to the angle between the current and the DW, leading to a distortion of the initially circular bubble towards an elliptical shape. Second, when $H_{ip}$ is applied, the bubble not only shifts parallel to the current, it also shifts in the perpendicular direction.

These distorted bubble displacements stem from the specific angular dependence of the DW velocity with respect to the electric current[31–33]. To verify whether these observations are consistent with our understanding of the FIDM and CIDM, we use the 1D model to calculate the velocity as a function of this angle. The shifting of the bubbles (white arrows in Figure 4a) is consistent with the asymmetry of the DW velocity predicted by the model (black arrows in Figure 4b), only if $\alpha_c$ is included. This shows that $\alpha_c$ is fully responsible for the bubble distortion observed at large $H_{ip}$.

**IV) Model predictions**

Up to now, we have shown that all our experiments can be understood by including simultaneously the effects of both $\alpha_c$ and DMI. Moreover, the successful fitting of the entire dataset has allowed to establish numerical values for the different parameters (see Methods). However, one should be cautious when further using these values. They are not real values, but rather "effective values" which are model dependent in the sense that they incorporate all the approximations made in the model. For example, we neglect the second order uniaxial anisotropy, the even components of chiral damping[21] as well as the corrections to the gyromagnetic ratio[19,22], we use an effective DW width; the temperature variation of the parameters with Joule heating is not considered (saturation magnetization, uniaxial anisotropy, DMI, SOT etc.). Therefore, these values should not be used to assess the strength of the microscopic interactions causing $\alpha_c$ and DMI. Nevertheless, since the model is explaining well all our experiments, we can use it to distinguish the roles played in the dynamics by each individual parameter (Figure S5). Most importantly, we can also predict how these dynamics will evolve as these values change. In particular, we can focus on the range of values required for practical applications. From this perspective, there are two main requirements:

*i)* In order to stabilize the rigid Néel DW structure or small size SKs, the DMI has to be much stronger than in the case of our experiments. The ceiling value for $H_{DMI}$ is imposed by the spontaneous formation of helical phases, which obstructs the stabilization of single DWs or SKs[4].

*ii)* Efficient current induced motion requires larger DL-SOT compared to the Pt/Co/Pt layers used here, but not too large compared to $H_{DMI}$. In this sense, Figure 5a shows a calculation of DW velocity as a function of current density expressed as $H_{DL}/H_{DMI}$. The dependence can be approximately divided into two parts: in the low current regime, DW velocity increases with current, while in the second part, it reaches a plateau. The best efficiency of the CIDM is achieved at intermediate current ($H_{DL}/H_{DMI} < 0.5$) where the velocity is large, but not saturated.

The numerical calculations indicate that the influence of the chiral damping will become stronger, as the DL-SOT and DMI approach the optimal values for applications (Figure 5b, c). The angular dependence of the DW velocity (black curve in Figure 5b, c), will tend to distort the SK circular structure affecting its dynamical stability. The chiral damping will reduce the distortion and improve the stability (blue curve in Figure 5b, c). On the contrary, if $\alpha_2$ changes sign, it will accentuate the SK deformation (red curve in Figure 5b, c). To understand the effect of the deformation on the SK stability, we perform micromagnetic simulations (S10), including chiral damping. We observe (Figure 5e) that as long as the effect of current is insufficient to destabilize the SK structure, $\alpha_c$ modifies the SK velocity, but its influence on the anisotropic velocity is concealed by the rigidity of the magnetic structure. However, at larger driving force, close to the SK instability, the anisotropic velocity becomes relevant: $\alpha_c$ can either limit or enhance the SK distortion, influencing its stability. **NB**. The values of the damping coefficient used in the simulations are intentionally smaller than those of the 1D model. This is imposed by our simplified numerical integration of the chiral damping in the MuMax, which becomes approximate for larger damping. Moreover, the definition of the micromagnetic chiral component of the damping ($\alpha^{\mu}_c$) differs by a factor of 2 from its definition in the $q$-$\phi$ model (see Supplementary information S10).

Furthermore, the simulations of the SK trajectories (Figure 5d), allow to evidence the effects of $\alpha_c$ by order of importance:

*1st order*: The effects of damping can be observed in the linear response regime, where the driving force does not modify the chirality of the magnetic texture significantly. In this case, $\alpha_c$, can be approximated by a constant value that depends on the chirality (Figure 5d). Note that it is very different compared to the Gilbert damping, which does not depend on the chirality. In a given material, the Gilbert damping coefficient is the same, regardless of the type of dynamics (uniform mode precession, SK motion etc…), while in presence of $\alpha_c$ the effective damping depends on the chirality of the magnetic texture involved in the dynamics.

*2nd order*: As the driving force increases, the chiral texture distorts, and $\alpha_c$ creates *non-linear damping* (Figure 5e). The SK trajectories deviate significantly from those obtained using a constant value. In this case, the damping varies in time and space (along the SK perimeter), and the evolution of the micromagnetic structure cannot be reproduced using a constant damping.

In conclusion, by performing a full set of experiments of DW and SK dynamics in Pt/Co/Pt, we were able to observe how chiral dissipation influences the DW motion in the flow regime, and also to understand how it affects the distortion of skyrmionic bubbles. In all experiments, we observe a reversal of the chiral asymmetry when an in-plane field is applied. The numerical model reproduces this behavior only if both $\alpha_c$ and DMI are included. In this case, all experimental features are comprehensively reproduced using a unique set of physical parameters. By extrapolating our numerical model to the range of large SOT and strong DMI, which is more suitable for applications, we evidence a strong influence of $\alpha_c$ on the SK dynamics and stability. Understanding and controlling $\alpha_c$ is thus essential for engineering devices employing DWs or SKs.

**Methods**

**FIDM**: We prepare DW bubbles in the un-patterned film by applying a small perpendicular field (5 mT for 1 s). Then we apply nanosecond field pulses (up to 217 mT for 40 ns) and extract the velocity from the DW displacements. Because of its small size and of the short pulses, the magnetic field of the micro-coil cannot be measured directly. It is calibrated by comparing low field DW velocity produced by the micro-coil to low field DW velocity produced by a macroscopic, calibrated coil.

**CIDM**: In order to achieve large current densities, we pattern the samples into micrometer sized wires. The DWs are prepared using a perpendicular magnetic field. Nanosecond current pulses with current densities ranging from $0.1 \cdot 10^8$ A/cm$^2$ to $2 \cdot 10^8$ A/cm$^2$ are used to propagate the DWs in the micro-wires (Figure 2a). The velocity is determined at every current density from the linear dependence of the DW displacement vs pulse length.

When measuring the CIDM in presence of $H_{ip}$, we cannot use the same procedure as described above. In these experiments, the Joule heating together with $H_{ip}$ cause unwanted nucleations limiting the range of pulse amplitude and duration that we can use. Since the range of current pulses is reduced, we cannot quantify the DW velocity by the linear regression of the displacement vs. pulse length. Therefore, we have evidenced the chiral effects using the following procedure:

1. We set the current density to a value strong enough to produced DW motion for $H_{ip}= 0$, but not so strong to produce nucleations for the largest $H_{ip}$ (145 mT).
2. The exact pulse width and pulse sequence (number and frequency) are optimized for every $H_{ip}$ value in order to maximize the displacements, while avoiding nucleations.
3. As the current density remains constant, when $H_{ip}$ is decreased, the number and frequency of pulses have to be increased in order to maintain the size of the measured displacements. This is a consequence of the weaker driving force provoking more frequent DW pinning to defects.

Using this procedure, we cannot quantify the velocity value as precisely as we did for the CIDM without $H_{ip}$. For this reason, we compare the relative variation of the up/down and down/up DW displacements under identical conditions as we vary the chirality by $H_{ip}$.

**Numerical model:** The best agreement between the ensemble of experiments and the $q$-$\phi$ model was obtained with the following set of parameters: *DW width, $\Delta$=6 nm; $H_{DMI}$=30 mT; DW dipolar field $H_{Dip}$=30 mT; $\alpha_1$=0.4; $\alpha_2$=0.2.*

**Acknowledgements**

We thank J. Vogel and S. Pizzini for their help with the MOKE imaging and the micro-coil experiments, as well as for their critical reading and commenting on the manuscript. We acknowledge funding for this work from the European Research Council (ERC) under the European Union's Horizon 2020 research and innovation program (grant agreement No 638653 – Smart Design). MSG acknowledges funding from MRI-CNCS/UEFISCDI through grant PN-III-P4-ID-PCE-2020-1853.

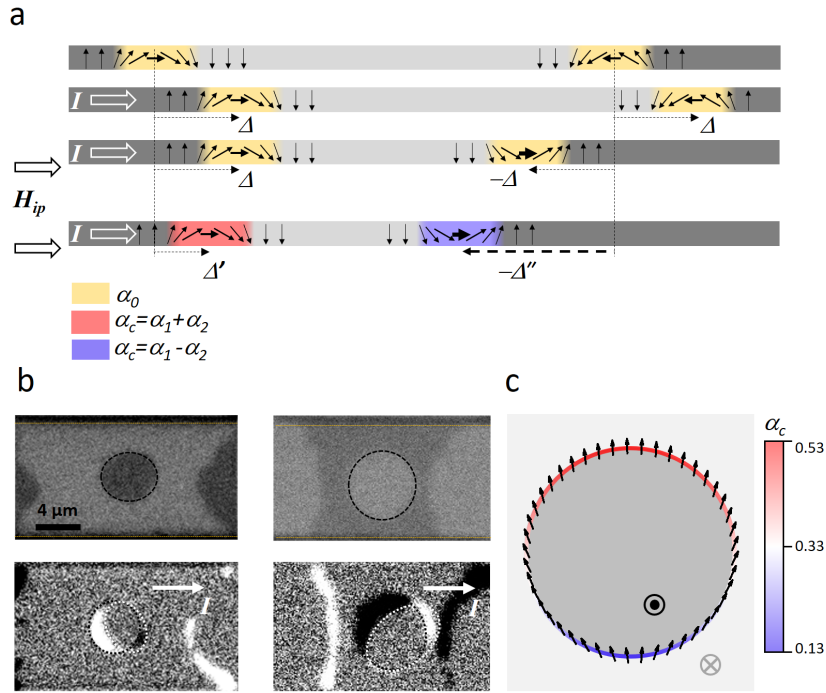

*Figure 1. Effect of chiral damping on the DW and SK dynamics* ***a.*** *Schematic representation of current induced motion of chiral DWs by SOT. The DWs with the same chirality (set by DMI) are displaced by the same amount (Δ represented by the dotted arrows) and in the same direction by the current induced SOT. When an external in-plane field sets the magnetization of the two DWs in the same direction, it imposes opposite chiralities. In this case, the two DWs still move by the same amount, but in opposite directions (Δ and −Δ). In the presence of chiral damping, the two DWs with different chirality experience different damping (red and blue). This engenders different mobility, which causes the different magnitude of the opposite displacements (**Δ′** and **−Δ″**).* ***b.*** *Kerr images of magnetic bubble motion driven by electric current pulses (5000 · 20 ns at $1·10^{12} A/cm^2$) in Pt/Co/Pt tri-layers. The two initial states correspond to bubbles magnetized either up (top-left) or down (top-right). The black dotted lines are guides for the eye indicating the initial bubble position. The orange dotted lines indicate the lateral edge of the magnetic pad. As seen in the differential Kerr images (below), the bubbles are displaced along the direction of the electric current (white arrow), but also undergo a chiral distortion (from circular to elliptical; the orientation of the ellipse depends on the chirality). The white dotted lines indicate the final state of the bubble.* ***c.*** *DW magnetization profile (black arrows) along the bubble perimeter under the effect of DMI and horizontally applied electric current ($H_{DMI}$ =200 mT; the damping-like and field-like components of SOT are $H_{FL}$=33 mT and $H_{DL}$=66 mT). The dark/bright grey indicates the up/down magnetization. This image is to illustrate that as the chirality changes along the perimeter of the bubble, the value of the chiral damping (color scale) will also vary, leading to an angular dependent DW mobility.*

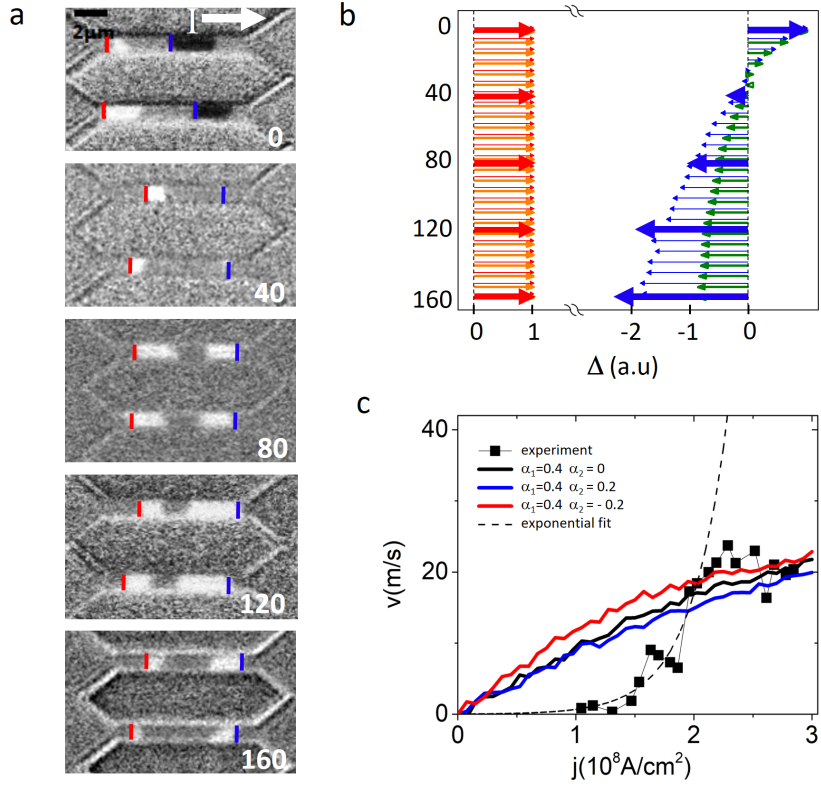

*Figure 2. Current induced DW motion a.* Differential Kerr images showing CIDM for different values of $H_{ip}$, shown in white on each image. Dark and white contrast indicates that up/down and down/up DWs are displaced in the same direction. Red and blue lines mark the starting positions for the DWs with opposite up/down polarities. The current density is fixed at $1·10^8 A/cm^2$. The pulse sequences used for each image are, from top to bottom:  10000 · 2 ns; 1000 · 1.5 ns; 100 · 1.5 ns; 200 · 1.2 ns; 300 · 1 ns **b.** Comparison of relative DW displacements extracted from the MOKE images (bold red and blue arrows) with numerical simulations. Since for every value of $H_{ip}$ in the experiment we had to use a different pulse sequence, it is impossible to extract accurately the DW velocity. Therefore, to evidence chiral effects, we compare the respective displacement of the two DWs by normalizing to the value of the positive displacement (in red). Orange and green arrows correspond to the scenario of $\alpha_1=0.4$ and $\alpha_2 = 0$. Red and blue arrows correspond to $\alpha_1=0.4$ and $\alpha_2 = 0.2$. In the absence of chiral damping, the displacements at large $H_{ip}$ become symmetric. The DWs move by the same amount in opposite directions. The reversal of the asymmetry observed experimentally, with the backward moving DW displacing a larger distance (blue arrows), is only reproduced in our model if we include the chiral damping. **c.** DW velocity as a function of current density (black symbols) for $H_{ip}=0$. The numerical model reproduces well the data in the range of high current density, above $2·10^8 A/cm^2$ and 20 m/s, within the boundaries of the flow regime. The dotted line is an exponential fit of the velocity in the intermediate range of current from $1·10^8 A/cm^2$ to $2·10^8 A/cm^2$. It is a guide for the eye to visualize the separation of the flow regime from the thermally activated DW motion.

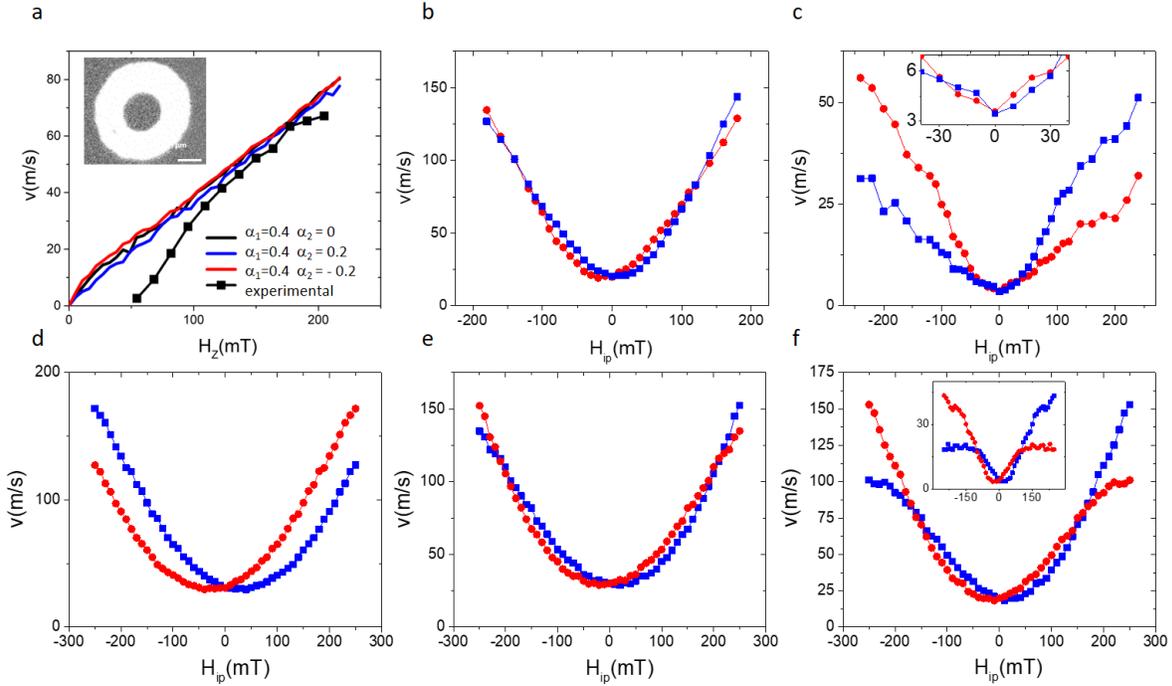

***Figure 3 Field induced DW motion a.*** *The experimentally measured DW velocity as a function of $H_Z$ (black symbols). The lines are the results of the model with DMI only (black) or including $\alpha_c$ (blue and red). The model reproduces accurately the velocity data in the flow regime (the region with constant DW mobility). The inset shows a typical differential Kerr image. The bright ring represents the DW displacement produced by the field pulse.* ***b.*** *Measured DW velocity vs. $H_{ip}$ (red for down/up DW, blue for up/down DW, see Figure S8 for detailed Kerr images) at constant $H_Z$= -82 mT and* ***c.*** *$H_Z$=-55 mT. The inset is a zoom of the low field region where the crossing of two curves occurs.* ***d, e, f,*** *Calculated DW velocity with (d) $H_Z$=-82 mT and $\alpha_c$=0.4; (e) $H_Z$=-82 mT and $\alpha_c$=0.4+0.2·$m_x$. (f) $H_Z$=-55 mT and inset $H_Z$=-10 mT*

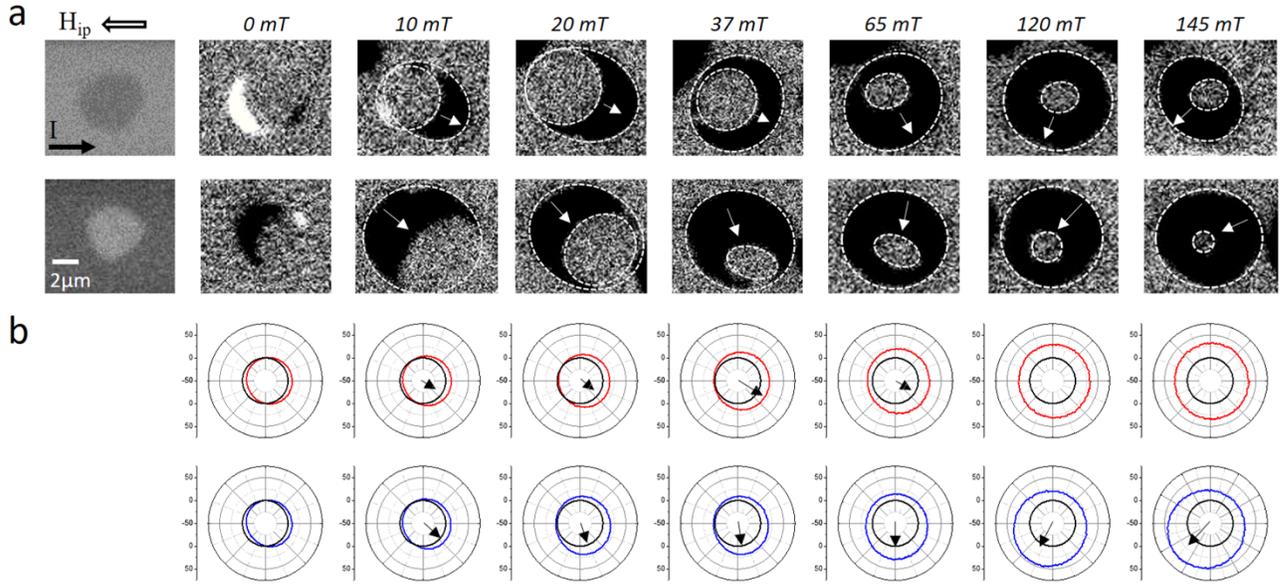

***Figure 4. Chiral distortion of skyrmionic bubbles***. **a.** *Kerr images of the bubble displacement under the effect of current pulses (20 ns at $1·10^8 A/cm^2$) and $H_{ip}$. On the first row we show images of a bubble magnetized up, on the second row, a bubble magnetized down. The first column shows the direct image of the initial bubble prepared by applying $H_Z$. The rest of the columns show differential images of displacements induced by the current for different values of $H_{ip}$ (from left to right: 0, 10, 20, 37, 65, 120, 145 mT). White dotted lines indicate the contours of the initial and final bubble. The white arrows point towards the direction of bubble displacement. The number of pulses is different for each image (top: 5000; 30000; 10000; 2000; 1500; 360; 120; bottom: 20000; 50000; 2500; 5000; 400; 100; 100;).* **b.** *Polar plot of computed DW velocity as a function of the angle between the DW orientation and electric current, corresponding to the same current density and $H_{ip}$ values as the experimental data. In the top row, we show in red results from a scenario with DMI but without $\alpha_c$. The bottom row shows in blue results including both $\alpha_c$ and DMI. From the velocity dependence, we extract the direction of bubble displacement indicated by the black arrows. They are in agreement with the white arrows on the MOKE images that indicate the experimental bubble displacement.*

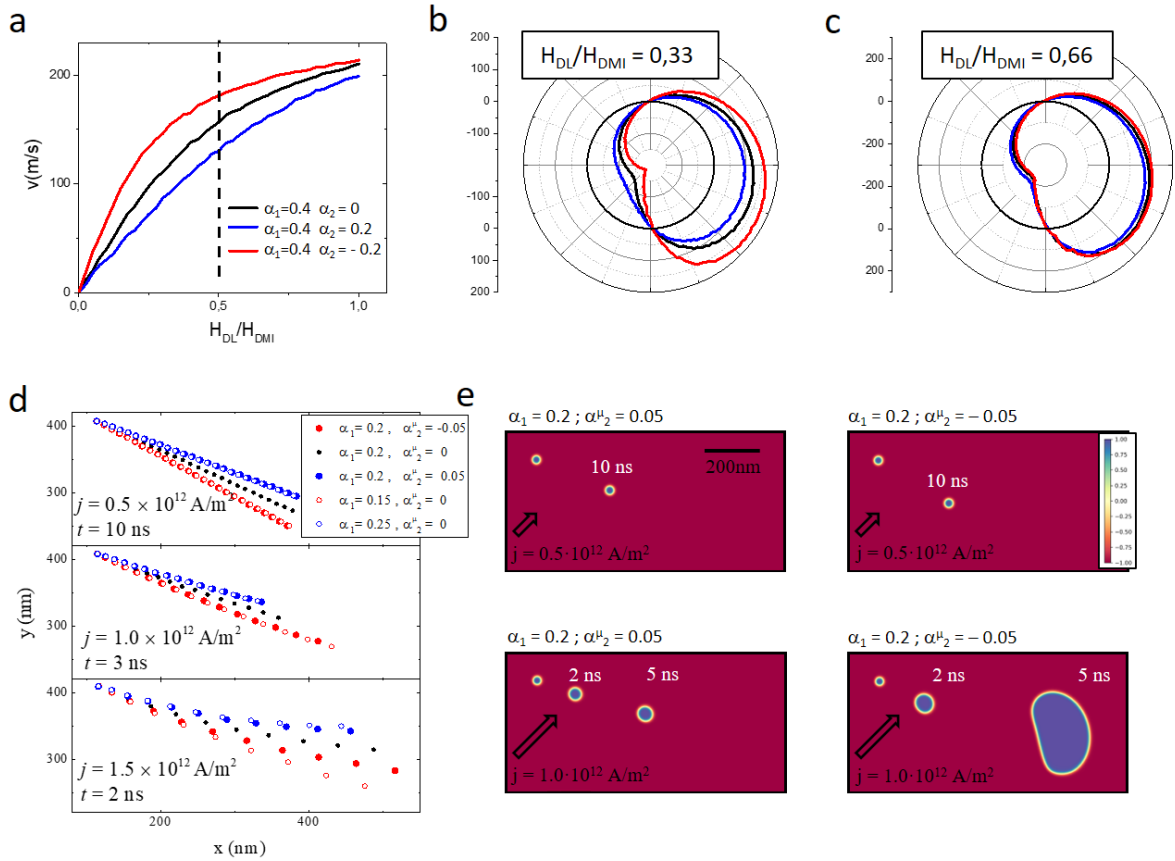

*Figure 5. Modelling of the SK distortion in presence of the chiral damping a.* CIDM plotted as a function of the ratio of $H_{DL}/H_{DMI}$. The dotted line approximately separates the linear variation regime from the plateau. Here, $H_{DMI}$= 0.2 T and $H_{DL}$=2·$H_{FL}$. The values chosen for these simulations are close to the optimal values and very importantly, they are realistic (typical for Pt/Co/AlO$_x$ systems). *b.* angular dependence of the velocity in the linear regime *c.* angular dependence of the velocity in the saturation regime. (black: $\alpha_1$=0.4; blue: $\alpha_1$=0.4, $\alpha_2$=0.2; red: $\alpha_1$=0.4, $\alpha_2$ = -0.2) *d.* Skyrmion position as a function of time for different current and damping values. *e.* Superposed images of the out-of-plane component of the micromagnetic structure at different moments of time during the simulation. On each image, the SK on the left corresponds to the initial state. At low current density (top row) the SK structure remains intact during the displacement. Chiral damping only influences the trajectory. At larger currents, at first the SK undergoes a slight distortion, which in the case of $\alpha_1$ = 0.2 and $\alpha^\mu_2$ = – 0.05, leads to its instability and the SK explodes. Here the electric current is applied at 45° with respect to the stripe major axis.

# Supplementary information

## S1. Numerical model

The DW motion at large driving forces is independent of material imperfections. For this reason, since the complexity of dealing with pinning and de-pinning effects is removed, this motion regime can be modelled more easily. Analytical models have been used successfully for materials with in-plane and out-of-plane anisotropy. Two motion regimes have been identified: steady DW motion, occurring at relatively low DW velocity, and turbulent motion at larger fields. In the steady regime, the DW structure is increasingly distorted as the velocity increases. Beyond a critical velocity, the DW internal structure is no longer stationary and transforms periodically.

In materials with perpendicular anisotropy, such as the ones used in the present study, the DW dynamics can be modeled by a few equations[1–3].

The field driven DW velocity in the steady regime:

$$v_{steady} = \frac{\gamma \Delta}{\alpha} H_z \qquad (1)$$

in the turbulent regime:

$$v_{turbulent} = \frac{\alpha \gamma \Delta}{1+\alpha^2} H_z \qquad (2)$$

The Walker critical field separating the two regimes:

$$H_W \approx \alpha \left( H_{DMI} + 1/2 \; H_{Dip} \right) \qquad (3)$$

The current driven DW velocity in the SOT-DMI model[4]:

$$v_j = \frac{\gamma \Delta H_{DMI}}{\sqrt{1+\left(C_o \cdot \frac{\alpha}{\Delta} \frac{H_{DMI}}{j \cdot \xi_{DL-SOT}}\right)^2}} \qquad (4)$$

With:

$$v_{turbulent} = \frac{\alpha \gamma \Delta}{1+\alpha^2} H_z \qquad (5)$$

Here, $\gamma$ is the gyromagnetic ratio; $\alpha$ is the damping; $\Delta$ is the DW width; $H_{DMI}$ is the DMI field; $H_{Dip}$ is the dipolar field associated to the Néel DW structure; $C_0$ is a constant depending on physical constants and sample parameters; $j$ is the current density; $\xi_{DL-SOT}$ is the efficiency of the damping like component of the SOT; $H_{ip}$ is the external in-plane field, $H_K$ is the anisotropy field of the sample. We note that the dynamics are determined by a relatively small number of parameters.

For the modelling of the DW motion, due to the complexity of our experiments, which include both magnetic fields and electric current, applied simultaneously at different angles, the numerical modelling is more convenient than analytical calculations. For this, we have chosen to model the DW motion using a collective coordinates $q$-$\phi$ approach[7]. This has the advantage of providing accurate results for cases where analytical solutions may be too complex. To ensure the validity of our numerical model, we have tested it for the simpler cases with well-known analytical results described by the above equations.

Such numerical models have been widely used to study the DW dynamics. Their advantage over micromagnetic simulations is that the computation time is much faster. At the same time, they yield the same physical results in the vast majority of situations[7]. The reason for this is that the approximation that they rely on, that of a rigid DW as a solitonic quasiparticle, is sufficient in most instances.

Moreover, since they rely on effective parameters, the results can be interpreted easily, and the influence of each physical parameter can be identified directly.

Our numerical approach includes both the chiral energy and chiral damping.

Chiral energy appears as an effective magnetic field acting on the core magnetization of the DW.

It has been shown theoretically that the chiral damping can take different forms: it can have a high-order dependence[8] on $m_x$ and it can even affect the gyromagnetic ratio[5]. Our model only uses the simplest form ($\alpha_c = \alpha_1 + \alpha_2 \cdot m_x$), which is sufficient to account for all the chiral features seen in our experiments. We note that while higher order contributions may exist and can affect the dynamics, disentangling the second order effects, or the different forms of chiral damping is beyond the purpose of our present work.

Non-chiral parameters such as the current induced torques, the DW dipolar field, domain wall width are included. The effect of disorder is included in the form of temperature fluctuations.

The current induced torques as well as the anisotropy field were determined experimentally using independent (second harmonic) measurements. $\xi_{DL-SOT} = 0.017$ (T/$10^{12}$ A/m$^2$), $\xi_{FL-SOT} = 0.004$ (T/$10^{12}$ A/m$^2$).

The values that fit best our results ($\Delta = 6$ nm; $H_{DMI} = 30$ mT; $H_{Dip} = 30$ mT; $\alpha_c = 0.4 + 0.2 \cdot m_x$) are uniquely determined and allow to model simultaneously the entire range of different experiments. The exact values of the non-chiral parameters affect little the general shape of the curves and have a weak influence on the emergence of chiral features (S6, S7).

**The fitting procedure**

The model is based on 5 independent fitting parameters, 2 fixed measured parameters:

Independent: $\alpha_1$, $\alpha_2$, $H_{DMI}$, $\Delta$, $H_{Dip}$.

Measured: $\xi_{DL-SOT}$, $\xi_{FL-SOT}$.

Equations (1) – (5) show that these are all the parameters that can influence the DW motion.

We fit the following experimental features:

- The value of the velocity saturation of the CIDM in the flow regime (Figure 2c)
- The value of the DW mobility in the flow regime of FIDM (Figure 3a)
- Reversal of the unidirectional component of the CIDM (Figure 2a, b)
- Reversal of the chiral asymmetry of the FIDM (Figure 3b, c, d, e, f)
- Asymmetry of the current-induced 2D bubble distortion produced by $H_{ip}$ (Figure 4)

From previous measurements in the creep regime we can apply more constraints on the chiral damping and DMI. From the magnitude and the saturation field of the anti-symmetric component of the DW motion in the creep regime[6], we had previously determined that

$\alpha_2/\alpha_1 \approx 0.5$ and

$H_{DMI} + H_{Dip} \approx 50$ mT.

When including these additional constraints, the total number of independent experimental features (7) exceeds the number of free parameters (5). This means that the values that we extract for these parameters are uniquely determined.

To determine their numerical values, we varied the values manually to improve the fit. After a few iterations, we converged toward a set of values that reproduce reasonably well all the different experiments simultaneously.

In order to illustrate the influence of each parameter on the dynamics, in the following we will show how the independent variations of each parameter affect the results.

## S2. Influence of $H_{DMI}$

$H_{DMI}$ field models the effect of the DMI interaction on the DW magnetization. In a $q$-$\phi$ model, this field either points in the direction of DW motion, or is opposed to it, thereby favoring the Néel type DW configuration. It has a drastic effect on current induced DW motion, as it tends to align the magnetization with the electric current, thus making the DL-SOT more efficient in producing DW displacements (Figure S1a). Moreover, by stabilizing the DW structure, it also increases the $H_W$ value (Figure S1b) for the field driven DW motion.

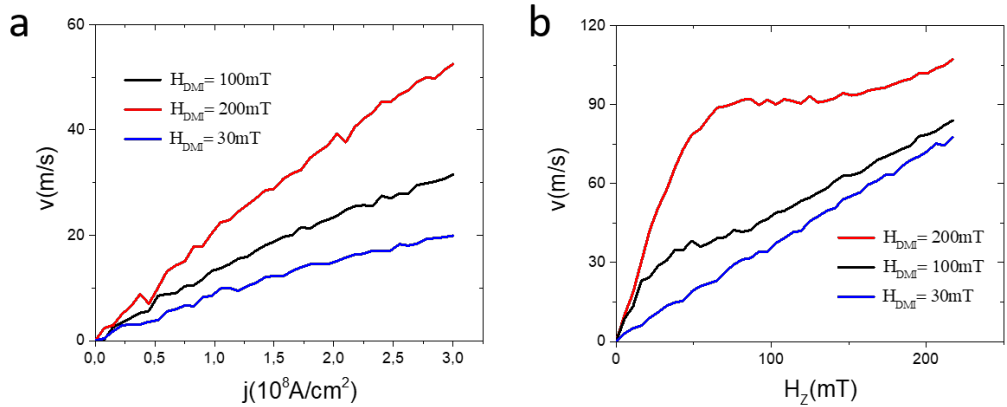

*Figure S1. a. Current induced DW velocity, b. field induced DW velocity, calculated for three different $H_{DMI}$ values.*

## S3. Influence of $\alpha_c$

For the field induced DW motion in the turbulent regime, there is a periodic change of the chirality, thus canceling the effects of the chiral dependent damping.

In the case of current induced motion, the insensitivity to damping is due to the fact that the ratio $H_{DL-SOT}/H_{DMI} > 1$. This is seen directly in equation (4): when the $H_{DL-SOT}$ is significantly larger than $H_{DMI}$, as it is in Pt/Co/Pt tri-layers, the velocity becomes less sensitive to the damping value[4].

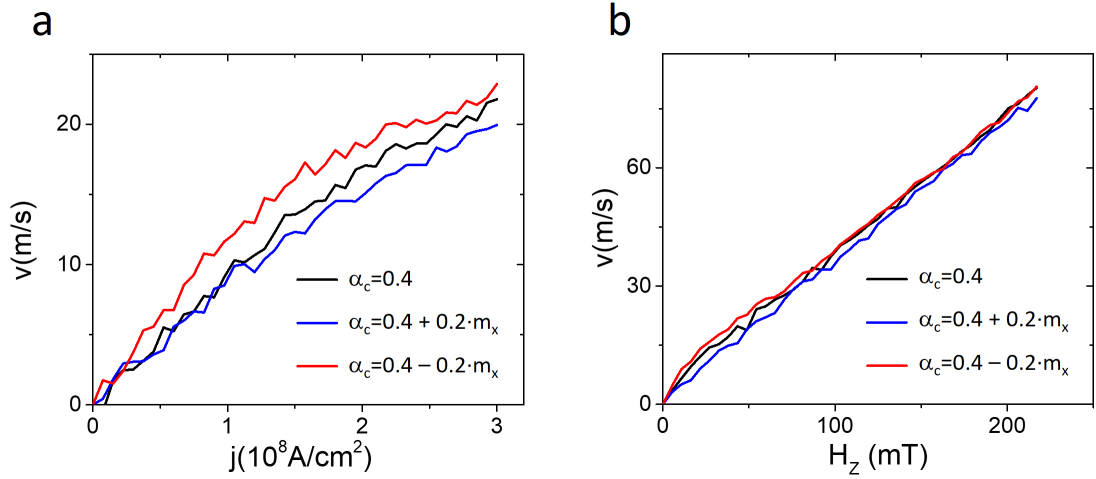

*Figure S2. a. Current induced DW velocity, b. Field induced DW velocity, for different $\alpha_c$ values*

### S4. Dipolar field of the DW

This effective field models the dipolar energy difference between the Bloch and Néel configurations. It will increase the stability of the Bloch DW structure and enhance the DW stability. Consequently, the increase of $H_{Dip}$ will tend to lower the efficiency of the CIDM and to increase the Walker breakdown field.

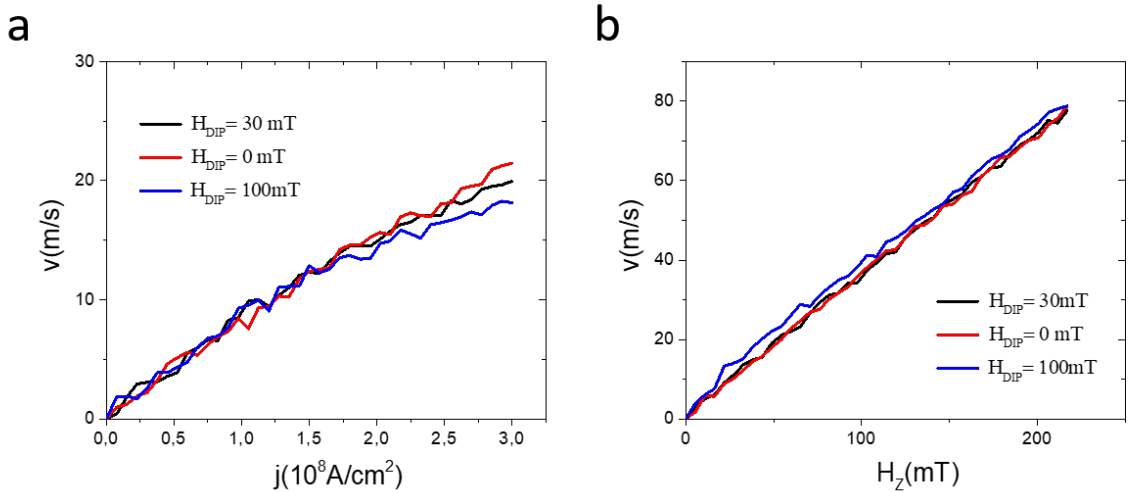

*Figure S3. a. Current induced DW velocity, b. Field induced DW motion, for different $H_{Dip}$*

### S5. Temperature

By increasing the disorder, the temperature lowers the $m_x$ component of the magnetization, thus reducing the efficiency of the CIDM. For the FIDM, the main effect of temperature is to smoothen the sharp features of the curves, such as Walker breakdown. Outside the range where the sharp features occur, the temperature value has little influence. The temperature is fixed at 300 K, while the "macrospin" volume subjected to temperature fluctuations is set to 0.5×6×10 nm for all the calculations. These dimensions correspond to the 0.5 nm = layer thickness, 6 nm= the DW width, and 10 nm = 2.5×$l_{ex}$ (the exchange length) an estimate of the length over which the thermal fluctuations do not disrupt significantly the parallel orientations of the spins[9].

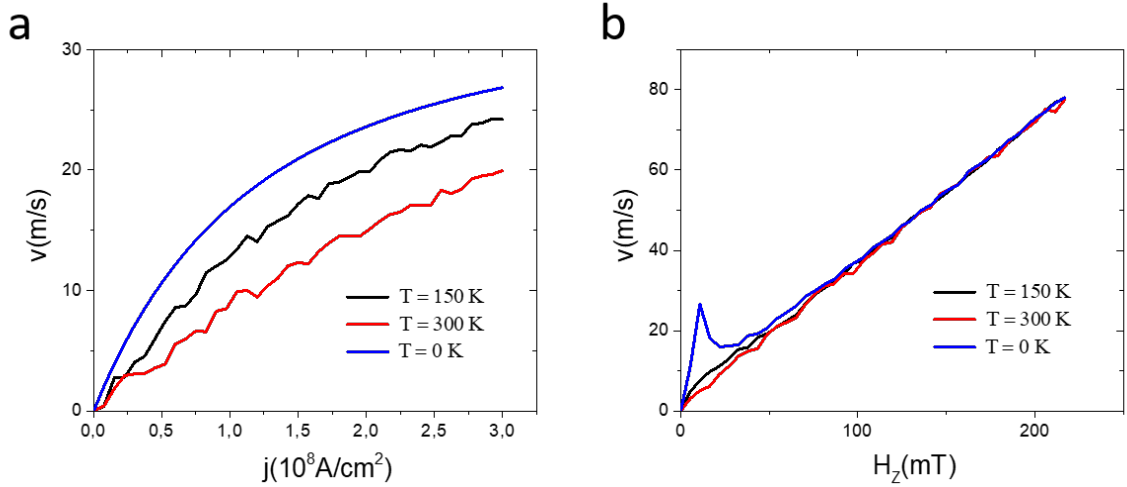

***Figure S4. a.*** *Current induced DW velocity,* ***b.*** *Field induced DW velocity for different temperatures*

## S6. Effect of $H_{ip}$ on the DW motion asymmetry

*a) FIDM*

First we analyse the field induced DW motion in absence of any chiral effects (Figure S6). $H_{ip}$ increases the DW velocity as the motion passes from turbulent to steady. The velociy has an even dependence on $H_{ip}$; the effects for the two DW polarities are identical. Moreover, $H_{ip}$ has no effect on the steady motion regime.

The main effect of DMI is to compete with $H_{ip}$, shifting the curves for the two polarities in opposite directions. It is only when we include the chiral damping that the velocity asymmetry changes sign in the steady regime. However, $\alpha_c$ alone can not produce a significant asymmetry in the turbulent regime. In order to reproduce the reversal of the asymmetry observed experimentally, it is required to include both $\alpha_c$ and DMI.

*b) CIDM*

In the absence of chiral effects, the velocity curves for the two DWs reflect the symmetry of the SOT: the DWs do not move at $H_{ip} = 0$; when $H_{ip} \neq 0$ the two DWs move in opposite directions with the same velocity. The main effect of DMI is to shift the curves in opposite directions. Consequently, the DWs move even without $H_{ip}$. However, when $H_{ip}$ becomes sufficiently strong, the DWs still end up moving with the same velocity in opposite directions. In order to obtain different DW velocities at large $H_{ip}$, we need to introduce $\alpha_c$. On the other hand, $\alpha_c$ alone cannot be responsible for the DW motion at zero field. Both $\alpha_c$ and DMI are required to reproduce the experimental observations.

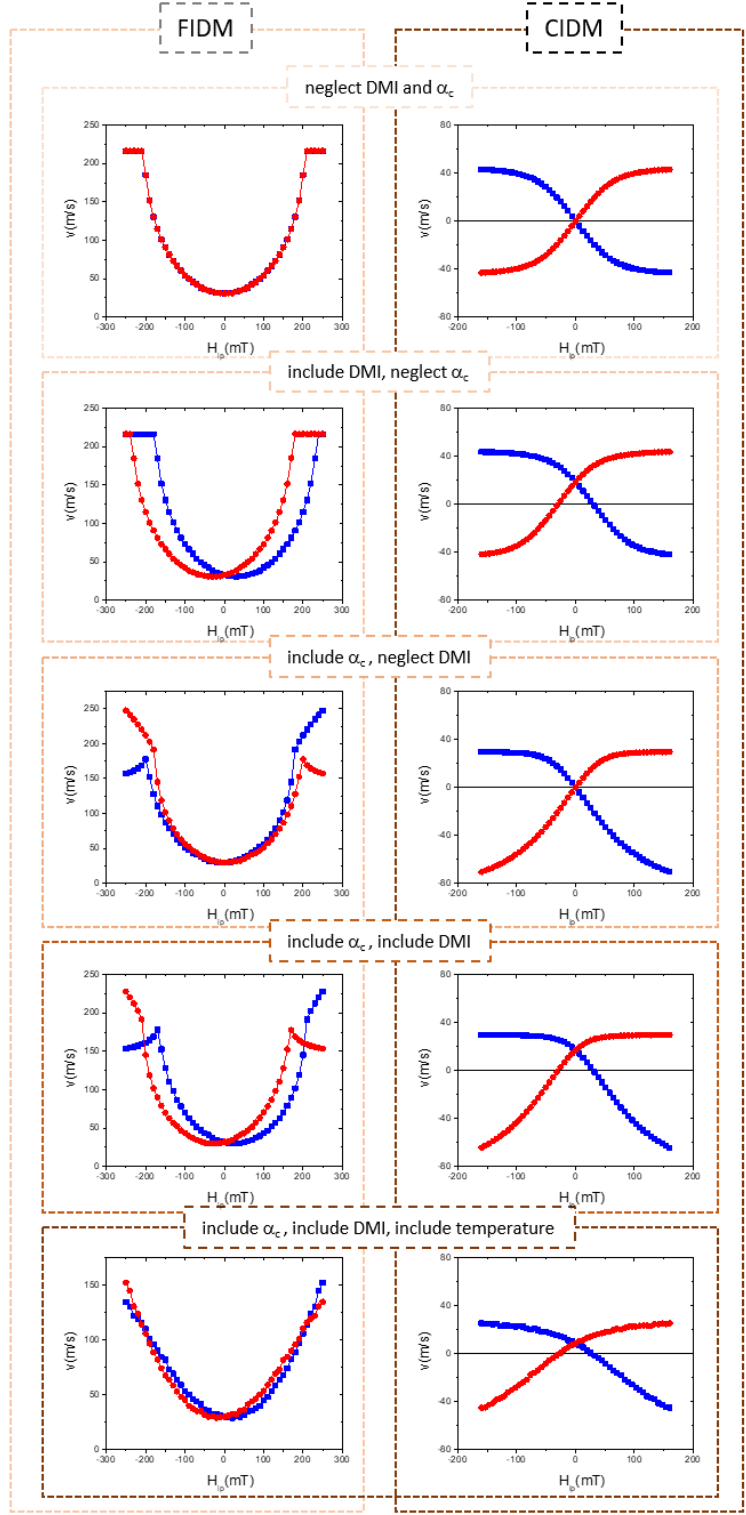

***Figure S5.*** *(left column) Effect of $H_{ip}$ on field driven motion ($H_Z$= 82 mT). We consider all the possible scenarios for the presence of chiral effects: no chiral damping and no DMI; only DMI; only chiral damping; both chiral damping and DMI. Temperature is also included. (right column) Effect of $H_{ip}$ on current driven motion ($j=10^8 A/cm^2$). We consider the same scenarios as for the field driven motion.*

## S7. Limit of the 1D model for 2D dynamics: the local approximation

We have extended the use of the 1D model to study the motion and distortion of the 2D bubbles under magnetic field and current. For this we have performed 1D calculations for all the angles of the DW canting with respect to the current and the applied field, and from here we have reconstructed the angular dependence of the velocity vs. DW angle. This approach is not exact, because the DW behaves similarly to an elastic membrane so that the behavior at a given angle is influenced by the DW behavior at adjacent positions.

For this reason, the measured data must be considered with care when comparing with models.

Nevertheless, experimentally we observe that the bubbles can have complex shapes (Figure S6); the shape depends on the magnetic history and is weakly influenced by the DW elastic energy. This can be understood when considering that the bubbles are relatively large and the pinning relatively strong. When the pinning energy overcomes the elasticity, the bubbles can undergo plastic deformation, and the different sections of the bubble evolve almost independently of the rest. In this case, the 1D approximation is sufficient to describe their deformation.

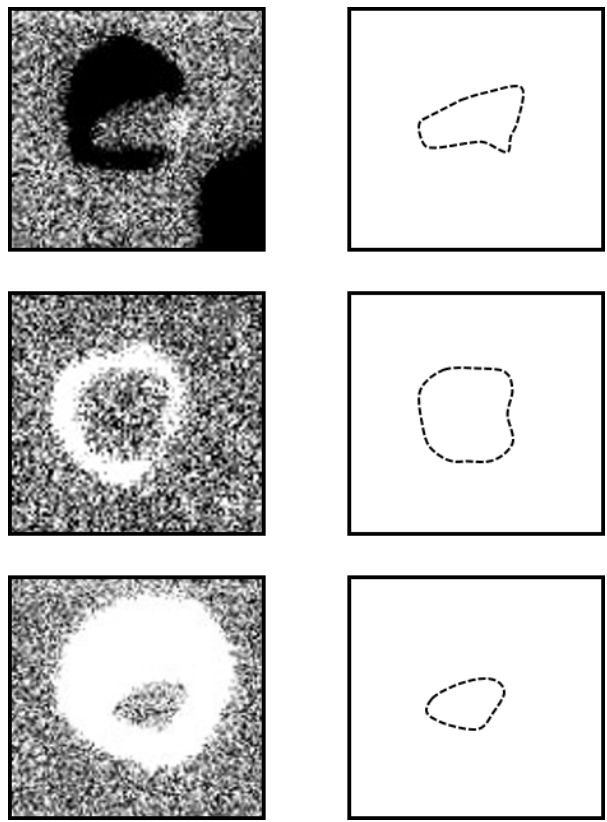

*Figure S6. Differential Kerr images of bubble growth. The shape of the initial bubble is extracted from each image and depicted on the right column. The bubble can have vastly different shapes, depending on*

*the precise sequence of magnetic field and current that was used to create it. This indicates an important plasticity of the bubble perimeter.*

## S8. MOKE images of CIDM in nanowires in presence of $H_{ip}$

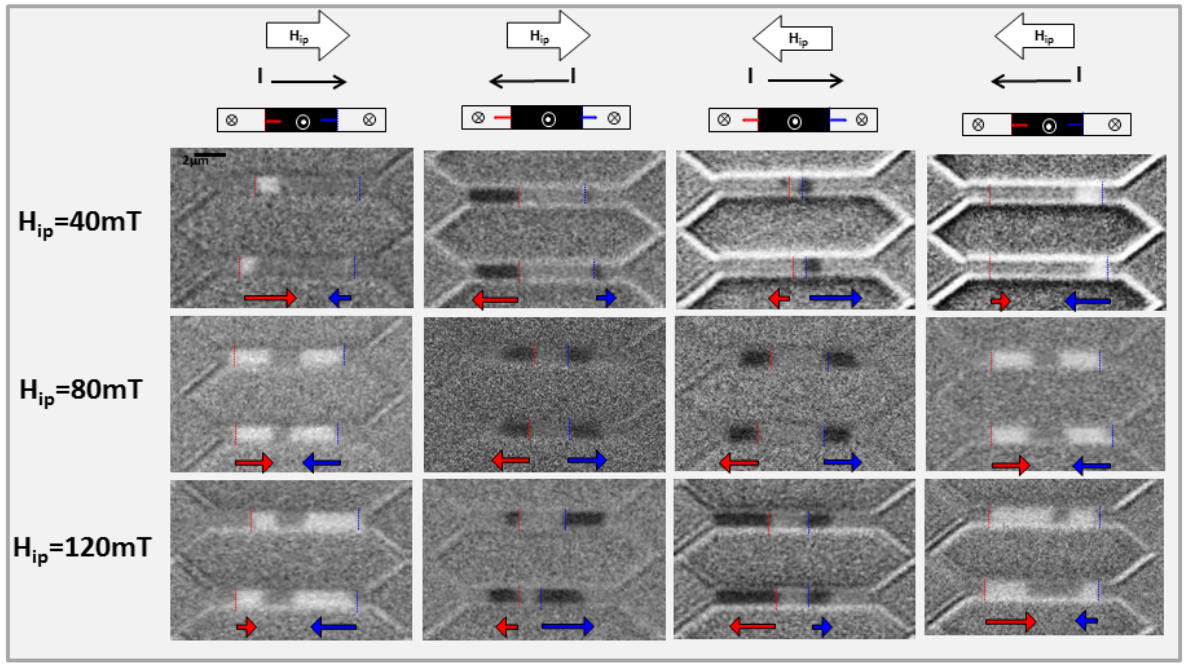

*Figure S7.* Differential Kerr images showing CIDM in the presence of $H_{ip}$= 40 mT, 80 mT, and 120 mT for 4 different combinations of current and $H_{ip}$ directions, depicted above the images. The motion of down/up and up/down DWs are schematically shown with red and blue arrows respectively. For each case, we clearly observe that the asymmetric DW motion at $H_{ip}$= 40 mT along one direction becomes almost symmetric at $H_{ip}$= 80 mT. At $H_{ip}$= 120 mT, the DW motion becomes asymmetric again, but the asymmetry direction is opposite to that observed at $H_{ip}$= 40 mT.

## S9. MOKE images of FIDM of a bubble in presence of $H_{ip}$

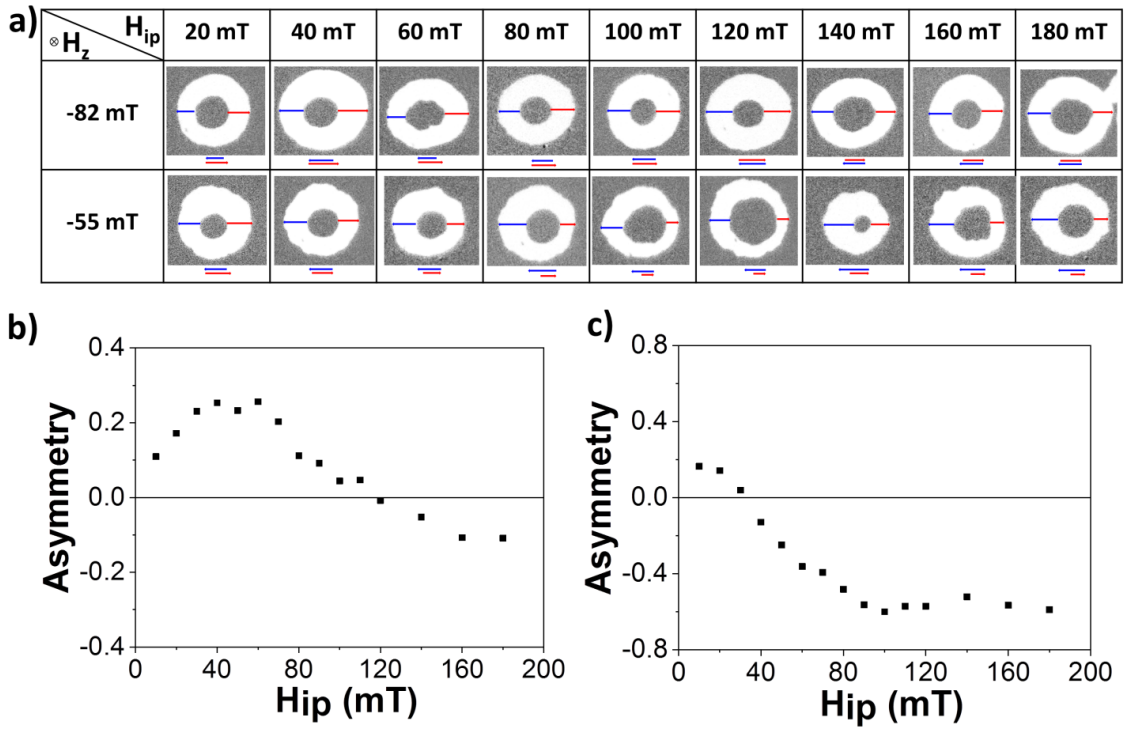

*Figure S8. a)* Differential Kerr images showing bubble expansion for $H_z$= -82 mT and -55 mT for different values of $H_{ip}$. The red and blue arrows schematically represent down/up and up/down DW displacements. The velocities extracted from these images were used in the plots shown in Figure 3 of the main text. Asymmetry= $\frac{v_{down/up}-v_{up/down}}{(v_{down/up}+v_{up/down})/2}$, calculated from the images *b)* at Hz= -82 mT *c)* at $H_z$= -55 mT. For both cases, the asymmetry direction reverses as $H_{ip}$ increases as explained in Figure 3 of the main text.

## S10. Micromagnetic simulations

The micromagnetic simulations were performed using the MuMax³ micromagnetic simulation software[10,11]. The chiral damping was implemented using the custom fields feature of MuMax³. As such, it is not included directly in the dissipative term, but rather as an effective field. The Landau–Lifshitz–Gilbert (LLG) governing the magnetization dynamics:

$$\frac{d\boldsymbol{m}}{dt} = -\gamma \boldsymbol{m} \times \boldsymbol{H}_{eff} + \alpha \boldsymbol{m} \times \frac{d\boldsymbol{m}}{dt}$$

can be transformed to:

$$\frac{d\boldsymbol{m}}{dt} = -\frac{\gamma}{1+\alpha^2} \boldsymbol{m} \times [\boldsymbol{H}_{eff} + \alpha(\boldsymbol{m} \times \boldsymbol{H}_{eff})]$$

where $\gamma$ is the gyromagnetic ratio and $\boldsymbol{H}_{eff} = -\frac{1}{\mu_0 M_s}\frac{\delta E}{\delta \boldsymbol{m}}$ is the effective field. Here, the energy includes the exchange, dipolar, anisotropy, DMI and SOT terms. Adding the chiral damping as an effective field is correct up to the renormalization of the gyromagnetic factor $(1+\alpha^2)^{-1}$. Using this approximate implementation, we cannot model exactly the experimental results in our samples, because the experimental damping is too large for this approximation to be applicable. Nevertheless, this correction becomes negligible when the damping is significantly smaller than one (so $\alpha^2$ becomes negligible relative to unity). For this reason, in the micromagnetic simulations we use small values of damping. The comparison to experiments is thus not quantitative, but qualitatively correct. Including the chiral damping the LLG reads:

$$\frac{d\boldsymbol{m}}{dt} = -\frac{\gamma}{1+\alpha_1^2} \boldsymbol{m} \times [\boldsymbol{H}_{eff} + \alpha_1(\boldsymbol{m} \times \boldsymbol{H}_{eff}) + \phi_C(\boldsymbol{m} \times \boldsymbol{H}_{eff})]$$

where the last term in the brackets represents the *chiral damping effective field* and $\phi_C$ is given by[4]:

$$\phi_C = \alpha^{\mu}{}_2 \Delta \left[ m_x \frac{\partial m_z}{\partial x} - m_z \frac{\partial m_x}{\partial x} + m_y \frac{\partial m_z}{\partial y} - m_z \frac{\partial m_y}{\partial y} \right]$$

where we introduce a characteristic exchange length $\Delta = \sqrt{\frac{A}{K_{eff}}}$ with $A$ the exchange constant and $K_{eff}$ the effective anisotropy, and $\alpha^{\mu}{}_2$ the chiral component of the damping coefficient. <u>Note that the micromagnetic definition of chiral damping is different from the $q$ - $\phi$ model definition. In the $q$ - $\phi$ model the chirality is expressed only with regard to the in-plane magnetization, while in the micromagnetic simulations the chirality is expressed both using the in-plane as well as the out-of-plane components. For this reason, equivalent results are obtained when the numerical value of chiral damping in the $q$ - $\phi$ model is approximately a factor of 2 larger than the value used in the micromagnetic simulations.</u>

For the simulations we used cells with a size of 1 nm. We chose realistic material parameters that were shown to stabilize skyrmions at room temperature in Pt/CoFeB/MgO layers[12]. Explicitly, the saturation magnetization $M_s = 1.12 \times 10^6 \, Am^{-1}$, perpendicular uniaxial anisotropy $K_{u1} = 10^6 \, Jm^{-3}$, exchange constant $A = 10 \times 10^{-12} \, Jm^{-1}$, the DMI constant $D = 1.5 \, mJm^{-2}$, the spin Hall angle $\theta_{SH} = 0.07$ and the ratio between the field-like and the damping like SOTs $\xi = 0.5$. During the SK simulations, an out-of-plane field of -10.4mT was applied.

Before studying the effect of chiral damping on the SK motion and their stability, we used the simulations to reproduce the main features of our experimental results. The goal is to validate the micromagnetic model that we will use further for studying the effect of chiral damping on the SK dynamics. It is to be noted that in this case we have used a lower DMI energy equivalent to a DMI field of $H_{DMI} = 10.4 \, mT$. The simulations reproduce well all the characteristic features observed experimentally and exhibit a very good qualitative agreement with our $q$-$\phi$ model (Figures S9 and S10 compared to Figure S5). This is an important point because the micromagnetic simulations require fewer assumptions compared to the $q$-$\phi$ model. For example, they automatically account for the changes of the DW structure (such as the variation of the DW width) during the application of the electric current and the magnetic fields. Furthermore, to test the robustness of our conclusions, we also used different values of DMI and chiral damping (Figure S9, Figure S10). They confirm that the competition between DMI and chiral damping asymmetries is a general effect, occurring throughout the entire parameter space; the asymmetry reversal only occurs when the chiral damping is sufficiently strong to overcome the asymmetry produced by DMI.

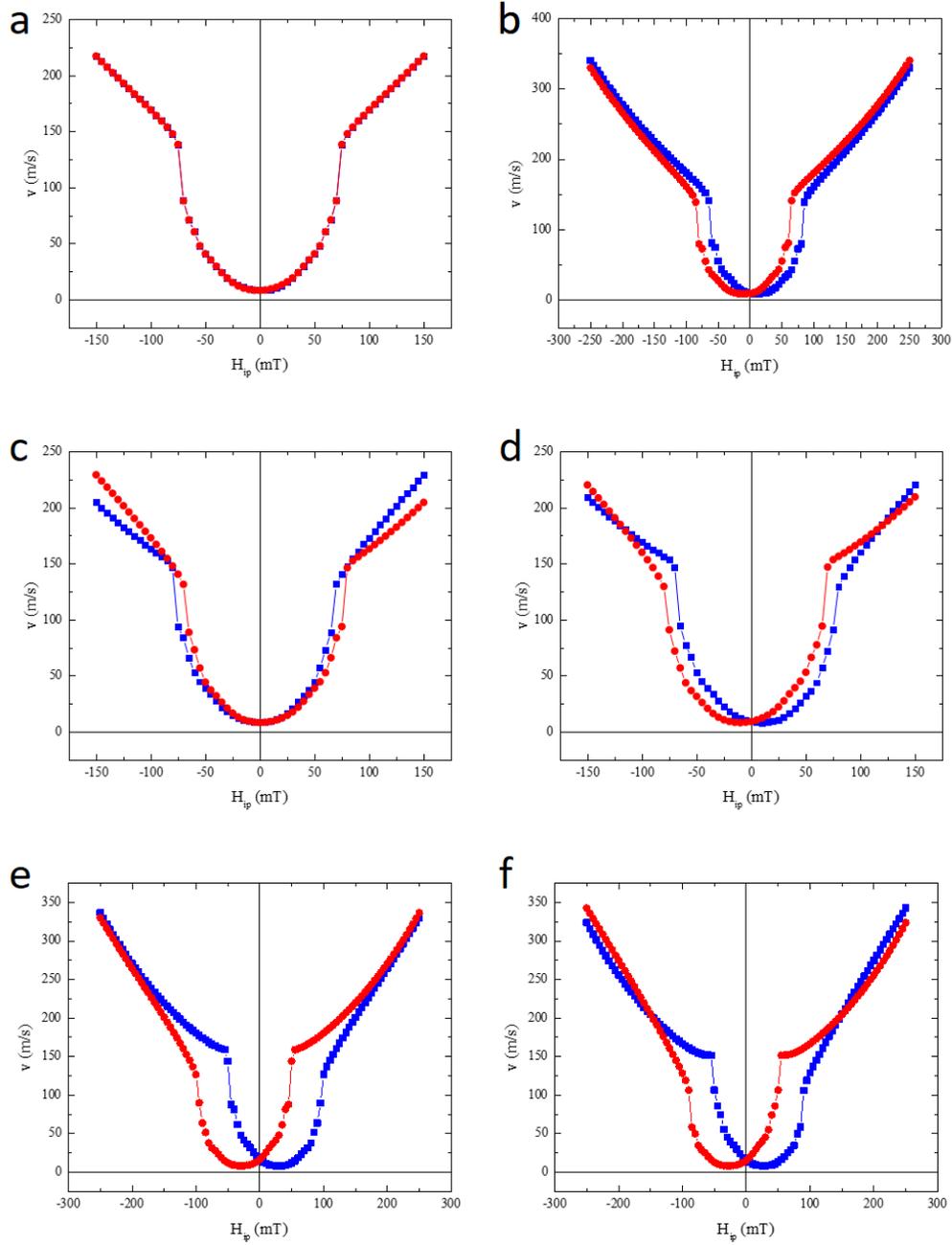

*Figure S9. Reversal of the asymmetry of the Field Induced DW motion ($H_Z$ = 30 mT). **a**. $H_{DMI}$ = 0, $\alpha_1$ = 0.2, $\alpha^\mu_2$ = 0. **b**. $H_{DMI}$ = 10.4 mT, $\alpha_1$ = 0.2, $\alpha^\mu_2$ = 0 **c**. $H_{DMI}$ = 0, $\alpha_1$ = 0.2, $\alpha^\mu_2$ = 0.05. **d**. $H_{DMI}$ = 10.4 mT, $\alpha_1$ = 0.2, $\alpha^\mu_2$ = 0.05. **e**. $H_{DMI}$ = 30 mT, $\alpha_1$ = 0.2, $\alpha^\mu_2$ = 0.05. **f**. $H_{DMI}$ = 30 mT, $\alpha_1$ = 0.2, $\alpha^\mu_2$ = 0.1*

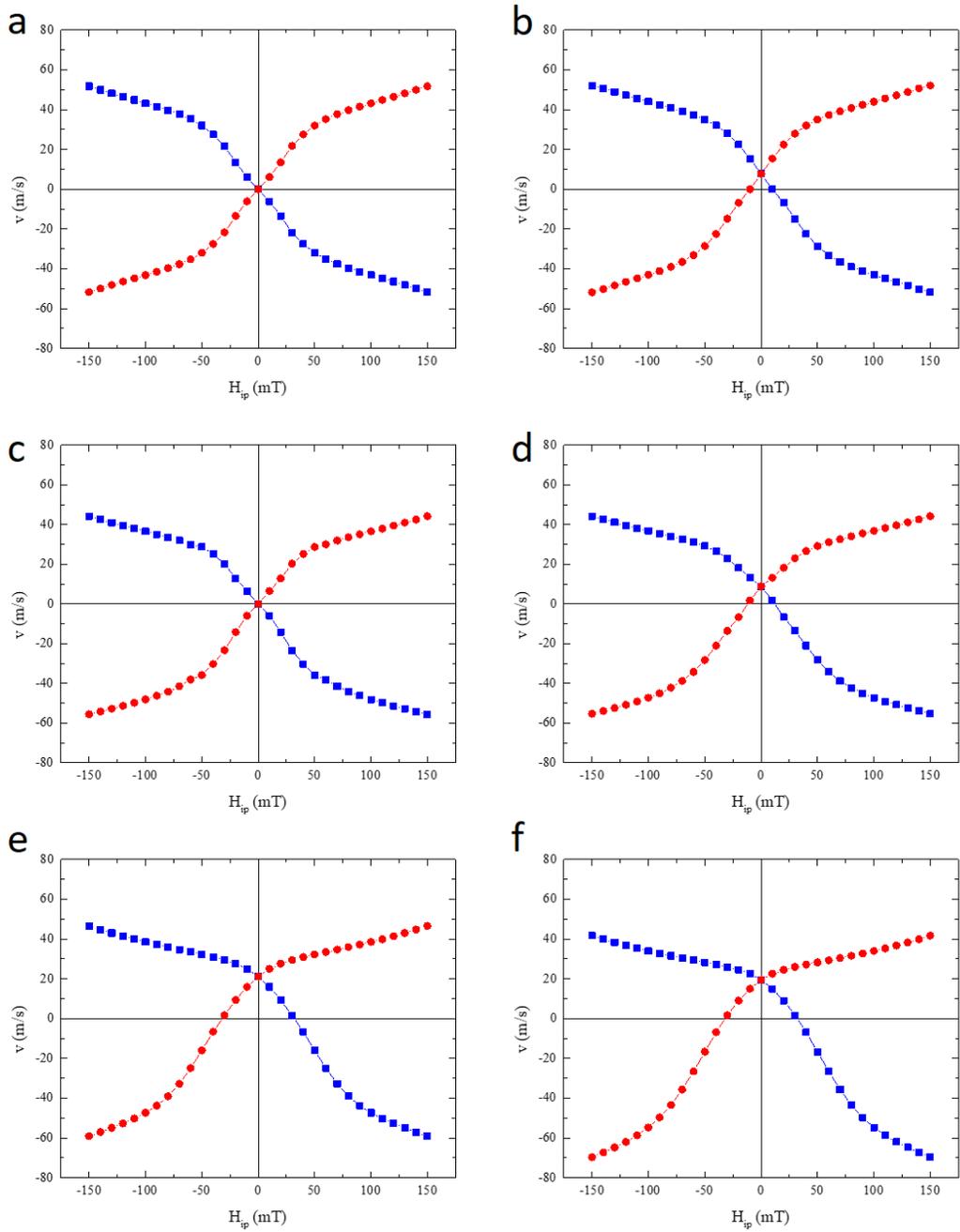

*Figure S10. Reversal of the asymmetry of the Current Induced DW motion ($j=2\times10^{11} A/m^2$)* **a**. $H_{DMI} = 0$, $\alpha_1 = 0.2$, $\alpha^u_2 = 0$. **b**. $H_{DMI}= 10.4$ mT, $\alpha_1 = 0.2$, $\alpha^u_2 = 0$ **c**. $H_{DMI} = 0$, $\alpha_1 = 0.2$, $\alpha^u_2 = 0.05$. **d**. $H_{DMI} = 10.4$ mT, $\alpha_1 = 0.2$, $\alpha^u_2 = 0.05$, **e**. $H_{DMI} = 30$ mT, $\alpha_1 = 0.2$, $\alpha^u_2 = 0.05$, **f**. $H_{DMI} = 30$ mT, $\alpha_1 = 0.2$, $\alpha^u_2 = 0.1$

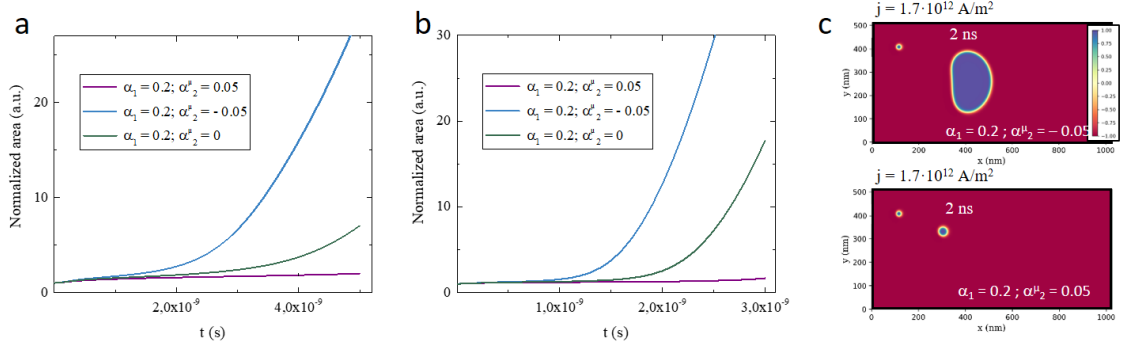

*Figure S11. Effect of the chiral damping on the Skyrmion dynamic stability. **a**. Evolution of the area of a dipolar stabilized skyrmion (seen in figure 5 of the main text) as a function of time for different chiral damping scenarios at $1.0 \times 10^{12}$ A/m². **b**. Area of a smaller, skyrmion stabilized by DMI (seen in figure 5 of the main text) as a function of time at $1.7 \times 10^{12}$ A/m². **c**. Micromagnetic simulation of the deformation of the small topological skyrmion for different chiral damping scenarios.*

In order to evidence the effect of the chiral damping on the SK stability we stabilize skyrmions with an equilibrium diameter of 40 nm and we simulate the effect of the SOT as a function of the current density for different values of damping.

The first effect of chiral damping is to provoke a variation of the SK velocity. Second, the damping influences the distortion of the skyrmion's circular shape. This is in agreement with the 2D velocity charts (Figure 5) calculated using the $q$-$\phi$ model.

The magnitude of the distortion is established by the competing effect of the asymmetric DW velocity and SK energy. If the SK is not sufficiently stable or the velocity asymmetry is too large, the SK size can diverge. In order to quantify this effect we use the total SK size as an indicator of the stability. In Figure S11 we plot the time evolution of SK size for a "large" dipolar stabilized SK as well as for a "small" SK stabilized by DMI. The majority of experimental studies realized at room temperature until now use the dipolar stabilized SKs. However, smaller skyrmions can be stabilized by using their topological protection. In our simulations, we were able to produce such "small" skyrmions, by slightly increasing the exchange coupling ($A = 12 \times 10^{-12}\ Jm^{-1}$). We confirm that in this case too, the chiral damping has a strong influence on the stability (Figure S11).